\documentclass[twocolumn,english,reprint]{revtex4-1}
\usepackage[T1]{fontenc}
\usepackage[latin9]{inputenc}
\usepackage{amsmath}
\usepackage{amssymb}
\usepackage{graphicx}
\usepackage{esint}

\makeatletter
\usepackage{adjustbox}

\makeatother

\usepackage{babel}
\begin{document}

\preprint{This line only printed with preprint option}

\title{Landau Damping of Electrostatic Waves in Arbitrarily Degenerate Quantum
Plasmas}

\author{Shane Rightley}

\email{shane.rightley@colorado.edu}

\selectlanguage{english}%

\author{Dmitri Uzdensky}

\email{uzdensky@colorado.edu}

\selectlanguage{english}%

\date{\today}

\affiliation{Center for Integrated Plasma Studies, Physics Department, University
of Colorado, UCB 390, Boulder, CO 80309-0390, USA}
\begin{abstract}
We carry out a systematic study of the dispersion relation for linear
electrostatic waves in an arbitrarily degenerate quantum electron
plasma. We solve for the complex frequency spectrum for arbitrary
values of wavenumber $k$ and level of degeneracy $\mu$. Our finding
is that for large $k$ and high $\mu$ the real part of the frequency
$\omega_{r}$ grows linearly with $k$ and scales with $\mu$ only
because of the scaling of the Fermi energy. In this regime the relative
Landau damping rate $\gamma/\omega_{r}$ becomes independent of $k$
and varies inversly with $\mu$. Thus, damping is weak but finite
at moderate levels of degeneracy for short wavelengths.
\end{abstract}
\maketitle

\paragraph{Introduction.}Although the field of plasma physics has
traditionally been concerned with systems of classical particles,
the treatment of collective interactions of charged quantum particles
has been of interest for many years \citep{0038-5670-3-1-R04,Pines1961,Pines1962,Lindhard_1954}.
More recently, interest in quantum plasma physics has increased due
to the possible importance of collective plasma effects in microelectronic
systems \citep{0295-5075-84-1-17006}; the experimental realization
of warm, dense matter \citep{Glenzer2007}; and in the description
of astrophysical compact objects~\citep{Uzdensky2014}.

Generally, quantum effects are expected to be significant for physics
at the smallest scales. For plasma physics, this means systems in
which microscopic scales such as the Debye length $\lambda_{D}\equiv\left(T/4\pi ne^{2}\right)^{1/2}$
and gyro-radius $\rho_{e}=m_{e}cv_{\perp}/eB$ are on the order of
or smaller than the de Broglie wavelength of the particles, $\lambda_{dB}\equiv\hbar/\left(2\pi m_{e}T\right)^{1/2}$.
The proper treatment of such microscopic scales requires kinetic theory.
For this reason, developing a kinetic theory of quantum plasma physics
is of special interest.

One classic textbook problem of classical kinetic plasma physics is
the propagation, dispersion, and collisionless Landau damping of linear
electrostatic electron waves in unmagnetized plasmas \citep{landau1946}.
This problem has proven useful in illustrating the effects of quantum
physics on plasma phenomena as well \citep{Melrose2010a,2010JPlPh..76....7E}.
Landau damping in the quantum case is also of interest as it is potentially
observable in experiments involving nano-scale electronics, plasmonic
devices and warm dense matter \citep{Tyshetskiy2012,0295-5075-84-1-17006,Glenzer2007,2015PPCF...57e4004M,PhysRev.124.1387}
and of importance in dense astrophysical plasmas \citep{Kumar2012}.
Plasmon dispersion and damping has been observed in solid-state plasmas
\citep{2007JPCM...19d6207T,1976PhRvB..13.2451G,1975PhRvL..34.1330R}.
On the theoretical front, the linear quantum longitudunal dispersion
relation has been studied extensively in the literature in a variety
of limiting cases, including: the completely degenerate limit with
phase velocities above \citep{2010JPlPh..76....7E} and near \citep{Vladimirov2011,1991ZhETF.100.1483K}
the Fermi velocity; at arbitrary phase velocities in one-dimensional
plasmas for several temperatures \citep{Bonitzetal}; and the high
phase velocity limit for arbitrarily degenerate quantum plasmas \citep{Melrose2010,Melrose2010a}
including the ion-acoustic mode \citep{Mushtaq2009}. In addition,
there have recently been studies of the onset of nonlinearity and
particle trapping in quantum Landau damping \citep{Daligault2014,Brodin2015}.
However, a systematic examination of the linear dispersion relation
for all wavelengths in arbitrarily degenerate plasmas has so far not
appeared in the literature. There is, for example, no study of short-wavelength
wave propagation or damping in moderately degenerate plasmas. For
this reason, a thorough study of the dispersion of linear electrostatic
waves of any wavelength in both partially and completely degenerate
quantum plasmas is called for.

In this paper, we study linear one-dimensional (1D) longitudinal waves
in an arbitrarily degenerate, unmagnetized electron gas, with a neutralizing
uniform ion background. We map out numerically the complex dispersion
relation over a broad range of wavenumbers $k$ and levels of degeneracy
$\mu/T$, $\mu$ being the chemical potential and $T$ the temperature.
We then use this map of the complex frequency $\omega=\omega_{r}-i\gamma$
to analyze the key differences between non-degenerate and degenerate
cases and discuss the possibility of Landau damping in the degenerate
limit. We find that for long wavelengths ~$\omega_{r}$ is approximately
constant and the damping rate $\gamma$ is exponentially suppressed
for all values of degeneracy. At short wavelengths however, both $\omega_{r}$
and $\gamma$ grow linearly with $k$ so that the relative Landau
damping rate $\gamma/\omega_{r}$ becomes independent of ~$k$. The
short-wavelength relative damping rate exceeds unity in the classical
case, but decreases as $\left(\mu/T\right)^{-1}$ as $\mu\rightarrow\infty$.
This implies that it may be important to account for both the presence
of electrostatic waves and their damping in the description of a real
partially degenerate plasma, depending on the specific level of degeneracy.
This additionally implies that electrostatic waves of arbitrarily
short wavelength are able to propagate in a degenerate plasma.

\paragraph{Theory of Linear Electrostatic Waves in a Quantum Plasma.}Quantum
kinetic plasma physics can be cast in a particularly convenient form
by utilizing the phase-space formulation of quantum mechanics \citep{Tyshetskiy2011}.
One defines the 1-body Wigner function $F\left(\mathbf{x},\mathbf{p};t\right)\equiv W\left(\hat{\rho}\right)=\int d^{3}\mathbf{y}\thinspace\text{e}^{-2i\mathbf{p}\cdot\mathbf{y}/\hbar}\left\langle \mathbf{x}+\mathbf{y}\left|\hat{\rho}\right|\mathbf{x}-\mathbf{y}\right\rangle $\,\citep{PhysRev.40.749},
where $\hat{\rho}\equiv\sum_{i}\alpha_{i}\left|\psi_{i}\right\rangle \left\langle \psi_{i}\right|$
is the density operator. The operation $W$, called the Wigner transformation,
defines the quantum phase space and $F$ acts as a quasi-distribution
function.

For a system with a scalar potential $V$ and Hamiltonian $H=p^{2}/2m+V$,
the Wigner function evolves according to the Moyal equation \citep{PSP:1593184,Vladimirov2011},\begin{widetext}

\begin{equation}
\frac{\partial F(\mathbf{x,p},t)}{\partial t}=-\frac{2}{\hbar}\thinspace\text{sin}\left[\frac{\hbar}{2}\left(\frac{\partial}{\partial\mathbf{x}_{F}}\cdot\frac{\partial}{\partial\mathbf{p}_{H}}-\frac{\partial}{\partial\mathbf{x}_{H}}\mathbf{\cdot\frac{\partial}{\partial p}}_{F}\right)\right]F(\mathbf{x,p,}t)H(\mathbf{x,p,}t),\label{eq:moyal_diff}
\end{equation}
\end{widetext}where the subscripts indicate the function upon which
a given derivative operates and the sine of differential operators
is defined by its power series. This equation becomes the Vlasov equation
in the classical limit $\hbar\rightarrow0$, with corrections of higher
order in $\hbar$ acting as scattering/non-phase-space-volume-preserving
terms.

Using Eq. (\ref{eq:moyal_diff}) linearized about a background equilibrium
distribution function $F_{0}$, one can obtain the longitudinal dielectric
function \citep{2010JPlPh..76....7E,0038-5670-3-1-R04,Lindhard_1954,Tyshetskiy2011},

\begin{equation}
\epsilon=1+\frac{\omega_{p}^{2}m_{e}}{2\hbar k^{2}}\int d^{3}v\frac{F_{0}\left(\mathbf{v}+\Delta\right)-F_{0}\left(\mathbf{v}-\Delta\right)}{\omega-\mathbf{v}\cdot\mathbf{k}},\label{eq:dielectric}
\end{equation}
where $\Delta\equiv\hbar\mathbf{k}/m_{e}$ and $\omega_{p}\equiv\left(4\pi ne^{2}/m_{e}\right)^{1/2}$
is the classical electron plasma frequency. We integrate Eq. \ref{eq:dielectric}
over the directions perpendicular to the wave propagation $\mathbf{k}$,
obtaining a 1D problem with a reduced equilibrium distribution function
of the velocity parallel to $\mathbf{k}$, $f_{0}\left(v_{\parallel}\right)\equiv\int d^{2}v_{\perp}F_{0}\left(\mathbf{v}\right)$.
The remaining integral is along a contour $C$ following the real
$v_{\parallel}$ axis and diverting into the negative $\text{Im}\left(v_{\parallel}\right)$
half-plane to encompass the singularity at $v_{\parallel}=\omega/k$.
This diversion must be done in such a way that the contour also avoids
any singularities in the function $f_{0}$.

The dielectric function in Eq. (\ref{eq:dielectric}) differs from
the classical case in two ways. First, the velocity derivative of
the distribution function is replaced with a finite difference in
the plasmon momentum $\hbar k$. Second, quantum mechanics can influence
the background distribution $F_{0}$, which for arbitrarily degenerate
plasmas in thermal equilibrium is a Fermi-Dirac distribution $F_{0}\left(\mathbf{v},\mu\right)\propto\left[1+\text{exp}\left(\mathbf{v}^{2}/2m_{e}T-\mu/T\right)\right]^{-1}$.
The corresponding reduced 1D distribution, defined by integrating
over the perpendicular directions, is
\begin{equation}
f_{0}\left(v_{\parallel},\mu\right)=\frac{N\left(\mu\right)}{v_{T}}\thinspace\textrm{Ln}\left(1+\textrm{e}^{-m_{e}v_{\parallel}^{2}/2T+\mu/T}\right),\label{eq:fermi}
\end{equation}
where $N\left(\mu\right)=\left[\sqrt{\pi}\textrm{Li}_{3/2}\left(-\textrm{e}^{\mu/T}\right)\right]^{-1}$
is the degeneracy-dependent normalization, thus normalizing $f_{0}$
to $1$, and $v_{T}=\left(T/m_{e}\right)^{1/2}$ is the classical
electron thermal velocity. Here $\text{Li}_{3/2}$ is the polylogarithm
function of order~$3/2$. The ratio $\mu/T$ determines the level
of degeneracy of the system: a Maxwellian is recovered in the limit
$\mu/T\rightarrow-\infty$ and a distribution with $f_{0}\left(v>v_{F}\right)\equiv0$
in the degenerate limit $\mu/T\rightarrow+\infty$, where $v_{F}=\left(2E_{F}/m_{e}\right)^{1/2}$
is the Fermi velocity with $E_{F}=\left(\hbar^{2}/2m_{e}\right)\left(2\pi n\right)^{2/3}$
the Fermi energy.

By making a transformation to nondimensional variables $x=v_{\parallel}/v_{T}$,
$k\rightarrow k\lambda_{D}$, and $\omega\rightarrow\omega/\omega_{p}$
in Eq. (\ref{eq:dielectric}) along with Eq. (\ref{eq:fermi}), we
obtain\begin{widetext}
\begin{equation}
\epsilon\left(\omega,k\right)=\begin{cases}
\begin{array}{c}
1-N\left(\xi\right)\frac{1}{Hk^{3}}\int_{-\infty}^{\infty}\,\textrm{Ln}\left[\frac{1+\xi\textrm{e}^{-\left(x+Hk\right)^{2}}}{1+\xi\textrm{e}^{-\left(x-Hk\right)^{2}}}\right]\frac{dx}{x-\omega/\sqrt{2}k}\\
1-N\left(\xi\right)\frac{1}{Hk^{3}}\left\{ \text{P}\int_{-\infty}^{\infty}\,\textrm{Ln}\left[\frac{1+\xi\textrm{e}^{-\left(x+Hk\right)^{2}}}{1+\xi\textrm{e}^{-\left(x-Hk\right)^{2}}}\right]\frac{dx}{x-\omega/\sqrt{2}k}+i\pi\textrm{Ln}\left[\frac{1+\xi\textrm{e}^{-\left(\omega/k+Hk\right)^{2}}}{1+\xi\textrm{e}^{-\left(\omega/k-Hk\right)^{2}}}\right]\right\} \\
1-N\left(\xi\right)\frac{1}{Hk^{3}}\left\{ \int_{-\infty}^{\infty}\,\textrm{Ln}\left[\frac{1+\xi\textrm{e}^{-\left(x+Hk\right)^{2}}}{1+\xi\textrm{e}^{-\left(x-Hk\right)^{2}}}\right]\frac{dx}{x-\omega/\sqrt{2}k}+2i\pi\textrm{Ln}\left[\frac{1+\xi\textrm{e}^{-\left(\omega/k+Hk\right)^{2}}}{1+\xi\textrm{e}^{-\left(\omega/k-Hk\right)^{2}}}\right]\right\} 
\end{array} & \begin{array}{c}
\text{Im}\left(\omega\right)>0\\
\text{Im}\left(\omega\right)=0\\
\text{Im}\left(\omega\right)<0
\end{array}\end{cases}.\label{eq:split-1-1-1-1}
\end{equation}
\end{widetext}Here P is the principal value, $\xi\equiv\text{exp}\left(\mu/T\right)$,
and $H\equiv\sqrt{\pi}\lambda_{dB}/2\lambda_{D}=8\pi\sqrt{2}\hbar\omega_{p}/T$
is the dimensionless quantum recoil parameter.

\begin{figure}[t]
\includegraphics{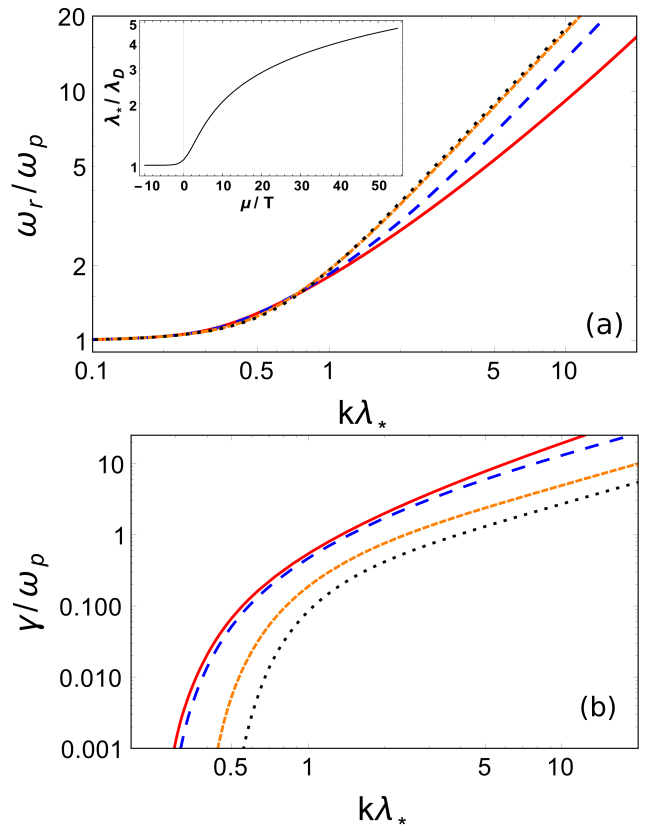}\caption{Dependence of the real (a) and imaginary (b) parts of the frequency
on wavenumber in units of the inverse generalized screening length,
$k\lambda_{*}\left(\mu\right)$, with $\mu/T$ fixed. Black (solid):
$\mu/T=-5$, blue (large dashes): $\mu/T=0$, orange (dashes): $\mu/T=5$,
red (dots): $\mu/T=10$\label{fig:real_kdep-1}.}
\end{figure}

In Eq. (\ref{eq:split-1-1-1-1}), there are three parameters: $k$,
$\xi$ (or $\mu$), and $H$. The important physical length and time
scales are $\lambda_{D}$ and $\omega_{p}^{-1}$. Additionally, there
exists a natural scale for wave phenomena given by the screening length
$\lambda_{*}\equiv\left\langle v\right\rangle /\omega_{p}$ where
$\left\langle v\right\rangle =\int\thinspace\left|\mathbf{v}\right|F_{0}\left(\mathbf{v},\mu/T\right)d^{3}\mathbf{v}$
is the average speed of particles in the plasma. The level of degeneracy
determines $\left\langle v\right\rangle $ and thus the screening
length is dependent upon degeneracy,
\begin{equation}
\lambda_{*}=\text{Li}_{5/2}\left(-\text{e}^{\mu/T}\right)\lambda_{D},
\end{equation}
which becomes the Debye length and Thomas-Fermi length $\lambda_{TF}=v_{F}/\omega_{p}$
in the non-degenerate and degenerate limits, respectively. The dependence
of this screening length on the chemical potential is shown in the
inset in Fig. \ref{fig:real_kdep-1}. All of the above parameters
and scales depend on $n$ and ~$T$.

The physical parameters of the system we describe are the density
and temperature $n$ and $T$, but these parameters are manifest in
the dielectric function Eq. (4) through the dimensionless parameters
$\mu\left(n,T\right)/T$ and $H\left(n,T\right)$. For simplicity
we have limited this study to the case of small $H=10^{-3}$, varying
only $\mu$ while keeping $H$ fixed. This means that $n$ and $T$
cannot take arbitrary values. In future work it will be beneficial
to map the dispersion relation for arbitrary values of $n$ and $T$
(and hence of $\mu$ and $H$), but for now we focus on the mathematical
properties of the dispersion relation without considering all possible
physical parameter regimes.

The quantum kinetic theory utilized in this paper is applicable for
weakly-coupled non-relativistic plasmas without spin. The first condition
requires a large number of particles to be present within a screening
radius $n\lambda_{*}^{3}\gg1$, and the second condition requires
that the thermal and Fermi velocities not approach the speed of light.
In the zero temperature (totally degenerate) case this has the effect
of setting minimum and maximum values for the density, namely $n\lambda_{TF}^{3}=\sqrt{\pi^{3}na_{B}^{3}}\gg1$
(where $a_{B}=\hbar^{2}/m_{e}e^{2}$ is the Bohr radius) and $v_{F}/c\ll1$.
Numerically this means $2.2\times10^{23}\text{cm}^{-3}\ll n\ll7\times10^{28}\text{cm}^{3}$
The extension to relativistic and spin plasmas is an ongoing effort.

\paragraph{Results.}We evaluate the dielectric function $\epsilon$
from Eq. (\ref{eq:split-1-1-1-1}) by direct numerical integration.
Taking the usual logarithm branch cut in the negative horizontal direction,
all integrals along the real axis are well-defined. We first solve
$\epsilon\left(\omega,k,\mu,H\right)=0$ for $\omega$ for single
values of $k$, $\mu/T$ and $H$ in the classical and $\omega/k\gg v_{T}$
limits using Newton's Method with a starting point given by an analytical
approximation. We then solve for the entire classical dispersion relation
by iterating to larger values of~$k$. We then iterate to larger values
of $\mu/T$ using the results of the next-smallest degeneracy. In
this way we map out $\omega$ in the entire $\left(k,\mu\right)$
plane with fixed value of $H=10^{-3}$.

\begin{figure}[t]
\includegraphics{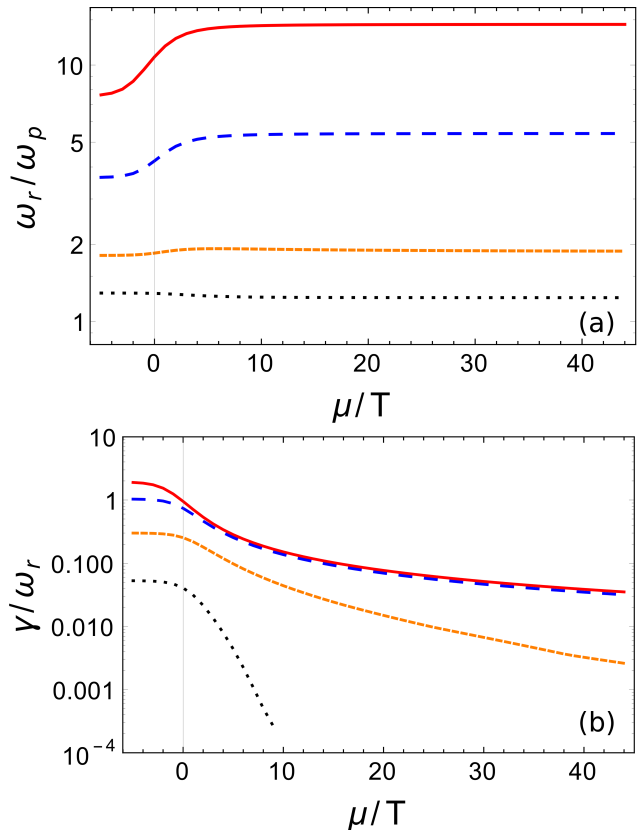}\caption{Dependence of $\omega_{r}$ (a) and the relative damping rate $\omega_{r}/\gamma$
(b) on level of degeneracy with $k\lambda_{*}$ fixed. Red (dotted):
$k\lambda_{*}=0.5$, orange (dashed): $k\lambda_{*}=1$ blue (large
dashes): $k\lambda_{*}=3$, black (solid): $k\lambda_{*}=8$.\label{fig:muvary}}
\end{figure}
~

A complication arises due to the dependence of the screening length
$\lambda_{*}$ on $\mu$, varying between $\lambda_{D}$ in the classical
case and $\lambda_{TF}\left(\mu\right)$ in the degenerate regime.
We thus consider the dependence of $\omega_{r}$ and $\gamma$ on
$\mu$ and $k\lambda_{*}$ instead of $k\lambda_{D}$. This dependence
is shown for several values of degeneracy in Fig. \ref{fig:real_kdep-1}.
In all cases, for small $k,$ $\omega_{r}$ is approximately $\omega_{p}$
and weakly dependent on $k$, and for $k\lambda_{*}>1$ grows linearly
with $k$. The damping rate $\gamma$ is exponentially suppressed
for small $k$ at all degeneracies, and also grows linearly at large
$k$. The effect of increasing degeneracy on $\omega\left(k\right)$
is to change the sharpness of the transition between flat and linear
growth of $\omega_{r}\left(k\right)$ near $k\lambda_{*}=1$, and
to shift $\gamma\left(k\right)$ to larger $k$.

The dependence of $\omega$ on $\mu/T$ for fixed $k\lambda_{*}$
is shown in Fig. \ref{fig:muvary}, and it is seen from examination
of $\omega_{r}$ that there are small- and large-$\mu$ regimes connected
by a transition region. This transition is caused by the change in
shape of the distribution function from Maxwellian to Fermi-Dirac.
The scaling of the distribution function with $\mu$ is thus accounted
for by the change in $\lambda_{*}$.

Since the primary goal of this paper is to understand the dependence
of Landau damping on the level of degeneracy, a quantity of particular
interest is $\Gamma\left(\mu\right)=\text{lim}_{k\rightarrow\infty}\gamma\left(k,\mu\right)/\omega_{r}\left(k,\mu\right)$.
We find that the quantity $\Gamma$ is approximately constant for
small $\mu/T$ and then turns downwards to decrease inversely with
degeneracy for larger $\mu/T$ approaching $\Gamma\left(\mu\right)\approx1.3\left(\mu/T\right)^{-1}$.

We have confirmed that the results obtained with our general numerical
method agree well with the previously published analytical results
in various limiting cases. For example, our calculations reproduce
the analytical approximations obtained by Melrose and Mushtaq \citep{Melrose2010a}
for the dependence of both $\omega_{r}$ and $\gamma$ on $k$ and
$\mu$ in the long-wavelength $k\lambda_{*}\ll1$ limit for arbitrary
degeneracy. Also, in the degenerate limit ($\mu/T\rightarrow\infty$)
our present results agree well with the solutions for $\omega_{r}$
derived for long and moderate wavelengths (small- and moderate $k$)
by \citep{1991ZhETF.100.1483K,2010JPlPh..76....7E} using a completely
degenerate top-hat equilibrium distribution with $F_{0}\left(v>v_{F}\right)\equiv0$.
For larger wavenumbers, Ref. ~\citep{1991ZhETF.100.1483K} claimed
that there is a critical maximum wavenumber above which no solutions
exist in the fully degenerate case, due to the non-analiticity of
the distribution function. Our present study (conducted for plasmas
with arbitrarily large, but finite degeneracy) cannot confirm this
result for the parameter regimes we have explored.

\paragraph{Discussion and Conclusions.}In conclusion, in this paper
we analyzed the complex frequency spectrum of linear longitudinal
electrostatic waves in an arbitrarily degenerate electron plasma with
a stationary neutralizing positive background. Using an appropriate
analytical form for the dispersion relation from quantum kinetic theory,
we developed a numerical procedure to solve for the real part of the
frequency $\omega_{r}$ and the Landau damping rate $\gamma$ as functions
of the wavenumber $k$ and the level of degeneracy $\mu/T$. We found
that there are two ways in which degeneracy affects the dispersion
relation: (1) through the change in the shape of the distribution
function from Maxwellian to a Fermi-Dirac profile, and (2) the increased
characteristic velocity as a function of~$\mu/T$. Above a certain
level of degeneracy, the distribution function assumes an approximately
constant shape; beyond this, the real part of the frequency only depends
on degeneracy through the scaling of the characteristic wavelength
$\lambda_{*}$. By accounting for this scaling we isolated the effect
of the shape of the Fermi-Dirac function on the complex frequency
$\omega=\omega_{r}-i\gamma$.

We showed that for wavelengths shorter than $\lambda_{*}$ both~$\omega_{r}$
and $\gamma$ grow linearly with $k$ for all levels of degeneracy
with the relative damping rate $\gamma/\omega_{r}$ decreasing as
$1.3\left(\mu/T\right)^{-1}$, independent of $k$ as $\mu/T\rightarrow\infty$.
This means that electrostatic waves in such systems propagate but
have finite damping rates. Since any real system occurs at finite
temperature, the choice of whether to account for the presence of
electrostatic waves, and whether to account for their damping, should
be informed by the specific temperature and density of a degenerate
system. This effect will be of importance not just for pure electrostatic
waves, but for other plasma waves in which collisionless kinetic damping
occurs.

A thorough understanding of Landau damping in the linear regime is
important before moving on to nonlinear effects. The realization that
nonlinear particle trapping is suppressed in certain parameter regimes
in quantum models \citep{Daligault2014} gives additional impetus
to study the linear regime of Landau damping, in contrast to classical
plasmas where the damping more quickly gives way to nonlinearity.
Additionally, the method used in this study can be easily extended
to further studies of linear waves and instabilities in quantum Fermi-Dirac
plasmas. Specifically, the inclusion of additional populations of
electrons or of mobile ions can be pursued in order to understand
streaming instabilities and ion-acoustic waves in quantum plasmas.
A study of these instabilities will be presented in a future publication.

\bibliographystyle{apsrev4-1}
\bibliography{refs}

\end{document}